\documentclass[final,copyright]{eptcs}
\providecommand{\event}{MBT-2015} % Name of the event you are submitting to

\usepackage[T1]{fontenc}
\usepackage[english]{babel}
\usepackage[utf8]{inputenc}
\usepackage{color}
\usepackage{amsthm}
\usepackage{amssymb}
\usepackage{amsmath,xspace}
\usepackage{cancel}
\usepackage[toc,page]{appendix}
\PassOptionsToPackage{normalem}{ulem}
\usepackage{ulem}
\usepackage{etex,etoolbox}

\makeatletter

\ifx \AtEndEnvironment \undefined
\newrobustcmd{\AtEndEnvironment}[1]{%
  \csgappto{@end@#1@hook}}

\patchcmd\end
  {\csname end#1\endcsname}
  {\csuse{@end@#1@hook}%
   \csname end#1\endcsname}
  {}
  {\etb@warning{%
     Patching '\string\end' failed!\MessageBreak
     '\string\AtEndEnvironment' will not work\@gobble}}
\else
\fi
\providecommand{\@fourthoffour}[4]{#4}
\def\fixstatement#1{%
  \AtEndEnvironment{#1}{%
    \xdef\pat@label{\expandafter\expandafter\expandafter
      \@fourthoffour\csname#1\endcsname\space\@currentlabel}}}

\globtoksblk\prooftoks{1000}
\newcounter{proofcount}

\long\def\proofatend#1\endproofatend{%
  \edef\next{\noexpand\begin{proof}[Proof of \pat@label]}%
  \toks\numexpr\prooftoks+\value{proofcount}\relax=\expandafter{\next#1\end{proof}}
  \stepcounter{proofcount}}

\def\printproofs{%
  \count@=\z@
  \loop
    \the\toks\numexpr\prooftoks+\count@\relax
    \ifnum\count@<\value{proofcount}%
    \advance\count@\@ne
  \repeat}
\makeatother
\theoremstyle{plain}
\newtheorem{thm}{\protect\theoremname}
  \theoremstyle{definition}
  \newtheorem{defn}[thm]{\protect\definitionname}
 \theoremstyle{definition}
 \newtheorem*{defn*}{\protect\definitionname}
  \theoremstyle{remark}
  \newtheorem*{rem*}{\protect\remarkname}
  \theoremstyle{plain}
  \newtheorem{lem}[thm]{\protect\lemmaname}
  \theoremstyle{definition}
  \newtheorem{example}[thm]{\protect\examplename}
  \theoremstyle{remark}
  
  \theoremstyle{plain}
  
  \fixstatement{thm}
\fixstatement{lem}

\usepackage{tikz}
\usepackage{caption}

\usepackage{subfig}
\makeatother

\usepackage{babel}
  \providecommand{\definitionname}{Definition}
  \providecommand{\examplename}{Example}
  \providecommand{\lemmaname}{Lemma}
  \providecommand{\propositionname}{Proposition}
  \providecommand{\remarkname}{Remark}
\providecommand{\theoremname}{Theorem}

\def\titlerunning{ioco theory for probabilistic automata}
\def\authorrunning{M.Gerhold, M.I.A. Stoelinga}

\newcommand{\pioco}{{\sf pioco}\xspace}
\newcommand{\ioco}{{\sf ioco}\xspace}

\begin{document}
\title{ioco theory for probabilistic automata}

\author{Marcus Gerhold \hspace{1.2cm} Mari\"elle Stoelinga
\institute{University of Twente, Enschede, The Netherlands}
\email{m.gerhold@utwente.nl \qquad marielle@cs.utwente.nl}
}

\maketitle

\begin{abstract}

Model-based testing (MBT) is an well-known technology, which allows for automatic test case generation, execution and evaluation. To test non-functional properties, a number of test MBT frameworks have been developed to test systems with real-time, continuous behaviour, symbolic data and quantitative system aspects. Notably, a lot of these frameworks are based on Tretmans' classical input/output conformance (ioco) framework. However, a model-based test theory handling probabilistic behaviour does not exist yet. Probability plays a role in many different systems: unreliable communication channels, randomized algorithms and communication protocols, service level agreements pinning down up-time percentages, etc.
Therefore, a probabilistic test theory is of great practical importance. We present the ingredients for a probabilistic variant of ioco and define the \pioco relation, show that it conservatively extends ioco and define the concepts of test case, execution and evaluation.

\end{abstract}

\section{Introduction}\label{sec(Introduction)}

Model-based testing (MBT) is a way to test systems more effectively and more efficiently. By generating, executing and evaluating test cases automatically from a formal requirements model, more tests can be executed at a lower cost.  A number of MBT tools have been developed, such as  the Axini test manager, JTorx \cite{DBLP:conf/tacas/Belinfante10}, STG \cite{STG}, TorXakis \cite{DBLP:conf/fmics/MostowskiPSTS09}, Uppaal-Tron \cite{DBLP:conf/fortest/HesselLMNPS08,LMNS05}, etc.

A wide variety of model-based test theories exist: the seminal theory of Input/Output conformance ~\cite{DBLP:journals/cn/Tretmans96,conf/fortest/Tretmans08} is able to test functional properties, and has established itself as the robust core with a wide number of extensions. The correct functioning of today's complex cyberphysical systems, depends not only on functional behaviour, but largely on non-functional, quantitative system aspects, such as real-time and performance.
MBT frameworks have been developed to support these aspects:
To test timing requirements, such as deadlines, a number of timed ioco-variants have been developed, such as  \cite{DBLP:journals/tcs/BensalemPQT08,DBLP:conf/fortest/HesselLMNPS08,DBLP:journals/fmsd/KrichenT09}. Symbolic data can be handled by the frameworks in \cite{FTW06,DBLP:journals/entcs/Jeron09}; resources by \cite{BS08}, and hybrid aspects in \cite{Osch06}.

This paper introduces \pioco, a conservative extension of ioco that is able to handle discrete probabilities. Starting point is a requirements model as a {\em probabilistic quiescent transition system (pQTS)}, an input/output transition system, with two additional features: (1) {\em Quiescence}, which models the absence of outputs explicitly via a distinct $\delta$ label: quiescence is an important notion in ioco, because a system-under-test (SUT) may fail a certain test case given an output is required, but the SUT does not provide one. (2) {\em Discrete probabilistic choice}. 
We work in the input-generative / output-reactive model \cite{GSST90}, which extend Segala's classical probabilistic automaton model \cite{Segala:1996:MVR:239648}: upon receiving an input, a pQTS chooses probabilistically, which target state to move to. For outputs, a pQTS chooses probabilistically both which action to take, and which state to move to, see Figure~\ref{fig:(ExampleGraph)} for an example.

An important contribution of our paper is the notion of test case execution and evaluation. In particular, we show how the use of statistical hypothesis testing can be exploited to determine the verdict of a test execution: if we execute a test case sufficiently many times and the observed trace frequencies do not coincide with the probabilities described in the specification pQTS depending on a predefined level of significance, then we fail the test case. In this way, we obtain a clean framework for test case generation, evaluation and execution. However, being a first step, we mainly establish the theoretical background. Further Research is needed to implement this theory into a working tool for probabilistic testing

\paragraph{Related work.}
An early and influential paper on probabilistic testing is {\em Bisimulation Through Probabilistic Testing} \cite{DBLP:conf/popl/LarsenS89}, which not only defines the fundamental concept
of probabilistic bisimulation, but also shows how different (i.e. non-bisimilar) probabilistic behaviours can be detected via statistical hypothesis testing.
This idea has been taken further in our earlier work \cite{CSV07,DBLP:conf/icalp/StoelingaV03}, which shows how to observe trace probabilities via hypothesis testing.

Testing probabilistic Finite State Machines is well-studied (e.g. \cite{Hwang20101108}) and coincidences to \ioco theory can be found. However pQTS are more expressive than PFSMs, as they support non-determinism and underspecification, which both play a fundamental role in testing practice. Hence, they provide more suitable models for today's highly concurrent and cyberphysical systems.

A paper that is similar in spirit to ours is by Hierons et al. \cite{hierons:hal-01055146,DBLP:journals/fac/HieronsN12}, and also considers input reactive / output generative systems with quiescence. However, there are a few important differences: Our model can be considered as an extension of \cite{hierons:hal-01055146} reconsiling probabilistic and nondeterministic choices in a fully fledged way.
Being more restrictive enables \cite{hierons:hal-01055146,DBLP:journals/fac/HieronsN12} to focus on individual traces, whereas we use trace distributions.

Other work that involves the use of probability is given in \cite{DBLP:conf/qsic/DulzZ03,DBLP:journals/tosem/WhittakerP93,DBLP:journals/infsof/WhittakerRT00}, which models the behaviour of the tester, rather than of the SUT as we do, via probabilities.

\paragraph{Organization of the paper.}
We start by defining overall preliminaries in Section \ref{sec(pqts)}.  
Section \ref{sec(pioco)} defines the conformance relation \pioco  for those systems and Section \ref{sec:Testing-for-pQTS} provides the structure for testing and denotes what it means for an implementation to fail or pass a test suite by the means of an output and a statistical verdict. 
The paper ends with conclusions and future work in Section~\ref{sec(FutureWork)}.

\section{Probabilistic quiescent transition systems}\label{sec(pqts)}

\subsection{Basic definitions}

\begin{defn}
(Probability Distribution) A \textit{discrete probability distribution}
over a set $X$ is a function $\mu:X\longrightarrow\left[0,1\right]$
such that $\sum_{x\in X}\mu\left(x\right)=1$. The set of all distributions
over $X$ is denoted as $\mathit{Distr}\left(X\right)$. The probability
distribution that assigns $1$ to a certain element $x\in X$ is called
the \textit{Dirac} distribution over $x$ and is denoted
 $\mathit{Dirac}\left(x\right)$.\end{defn}
\begin{defn}
(Probability Space) A \textit{probability space} is a triple $\left(\Omega,\mathcal{F},\mathbb{P}\right)$,
such that $\Omega$ is a set, $\mathcal{F}$ is a $\sigma$-field
of $\Omega$, and $\mathbb{P}:\mathcal{F}\rightarrow\left[0,1\right]$
a probability measure such that $\mathbb{P}\left(\Omega\right)=1$
and $\mathbb{P}\left(\bigcup_{i=0}^{\infty}A_{i}\right)=\sum_{i=0}^{\infty}\mathbb{P}\left(A_{i}\right)$
for $A_{i}$, $i=1,2,\ldots$ pairwise disjoint.\end{defn}

\subsection{Probabilistic quiescent transition systems}\label{subsec:pqts}

As stated, we consider probabilistic transitions that are \textit{input reactive} and \textit{output generative} \cite{GSST90}: upon receiving an input, the system decides probabilistically which next state to move to. However, the system cannot decide probabilistically which inputs to accept. For outputs, in contrast, a system may make a probabilistic choice over various output actions. 
This means that each transition in a pQTS either involves a single input action, and a probabilistic choice over the target states; or it makes a probabilistic choice over several output actions, together with their target states. We refer to Figure~\ref{fig:(ExampleGraph)} for an example. 

Moreover, we model quiescence  explicitly via a $\delta$-label. Quiescence means absence of outputs and is essential for testing: if the SUT does not provide any outputs, a test must determine whether or not this behaviour is correct. In the non-probabilistic case, this can be done either via the suspension automaton construction \cite{Tretmans96}, or via QTSs \cite{STS13}. The SA construction involves determinization. However, this is an ill-defined term for probabilistic systems. Therefore, we use the quiescent-labelling approach and demand to make quiescence explicit.

Finally, we assume that our pQTSs are finite and don't contain internal steps (i.e., $\tau$-transitions).

\begin{defn}
\label{def:(pQTS)}(pQTS) A \textit{probabilistic quiescent transition
system} (pQTS) is an ordered five tuple $\mathcal{A}=\left(S,s_{0},L_{I},L_{O}^{\delta},\Delta\right)$
where
\begin{itemize}
\item $S$ a finite set of states,
\item $s_{0}\in S$ the initial state, 
\item $L_{I}$ and $L_{O}^{\delta}$ disjoint sets of input and output actions,
with at least $\delta\in L_{O}^{\delta}$. We write $L:=L_{I}\cup L_{O}^{\delta}$
for the set of all labels and let $L_{O}=L_{O}^{\delta}\backslash\left\{ \delta\right\} $
the set of all real outputs.
\item $\Delta\subseteq S\times\mathit{Distr}\left(L\times S\right)$ a finite transition
relation such that for all  $\left(s,\mu\right)\in\Delta$,  $a?\in L_{I}$, $b\in L$,
$s',s''\in S$, 
 if $\mu\left(a?,s'\right)>0$, then $\mu\left(b,s''\right)=0$ 
for all  $b\neq a?$. 
\end{itemize}
We write $s\overset{\mu,a}{\rightarrow}s'$ 
if $\left(s,\mu\right)\in\Delta$ and $\mu\left(a,s'\right)>0$; 
and $s\rightarrow a$ if there are  $\mu\in\mathit{Distr}\left(L\times S\right)$ and $s'\in S$ such that  $s\overset{\mu,a}{\rightarrow}s'$. If it is not clear from the context about which system we are talking, we will write $s\overset{\mu,a}{\rightarrow}_\mathcal{A}s'$, $\left(s,\mu\right)_\mathcal{A}$ and $s\rightarrow_\mathcal{A} a$ to clarify ambiguities. Lastly we say that $\mathcal{A}$ is \textit{input enabled} if for all $s\in S$ we have  $s\rightarrow a?$ for every $a\in L_I$.
\end{defn}

\subsection{Paths and traces}

We define the usual language-theoretic concepts for pQTSs. 
\begin{defn}\label{defn(traces)}

Let $\mathcal{A}=\left(S,s_{0},L_{I},L_{O}^{\delta},\Delta\right)$
be a pQTS.
\begin{itemize}
\item A \textit{path} $\pi$ of a pQTS $\mathcal{A}$ is a (possibly)
infinite sequence of the form
\[
\pi=s_{1}\mu_{1}a_{1}s_{2}\mu_{2}a_{2}s_{3}\mu_{3}a_{3}s_{4}\ldots,
\]
where $s_{i}\in S$, $a_{i}\in L$ for $i=1,2,\ldots$ and $\mu\in\mathit{Distr}\left(L,S\right)$,
such that each finite path ends in a state and $s_{i}\overset{\mu_{i},a_{i}}{\rightarrow}s_{i+1}$
for each nonfinal $i$. We use the notation $\mathit{first}\left(\pi\right):=s_{1}$
to denote the first state of a path, as well as $\mathit{last}\left(\pi\right):=s_{n}$
for a finite path ending in $s_{n}$, and $\mathit{last}\left(\pi\right)=\infty$
for infinite paths. The set of all finite paths of a pQTS $\mathcal{A}$
is denoted by $\mathit{Path}^{*}\left(\mathcal{A}\right)$ and the set of all infinite paths by $\mathit{Path}\left(\mathcal{A}\right)$ respectively.

\item The \textit{trace} of a path $\pi=s_{1}\mu_{1}a_{1}s_{2}\mu_{2}a_{}s_{3}\ldots$
is the sequence obtained by omitting everything but the action labels, i.e. $\mathit{trace}\left(\pi\right)=a_{1}a_{2}a_{3}\ldots$.

\item All finite traces of $\mathcal{A}$ are summarized in $\mathit{traces}\left(\mathcal{A}\right)=\left\{ \mathit{trace}\left(\pi\right)\in L^{*}\mid\pi\in\mathit{Path}^{*}\left(\mathcal{A}\right)\right\} $.

\item We write $s_{1}\overset{\sigma}{\Rightarrow}s_{n}$ with $\sigma\in L^{*}$
for $s_{1},s_{n}\in S$ in case there is a path $\pi=s_{1}\mu_{1}a_{1}\ldots\mu_{n-1}a_{n-1} s_{n}$ with 
$\mathit{trace}\left(\pi\right)=\sigma$ and $s_{i}\overset{\mu_{i},a_{i}}{\rightarrow}s_{i+1}$ for $i=1,\ldots,n-1$. 

\item We write $\mathit{reach}{}_{\mathcal{A}}\left(S',\sigma\right)$ for
the set of reachable states of a subset $S'\subseteq S$ via $\sigma$,
i.e. \\
$\mathit{reach}_{\mathcal{A}}\left(S',\sigma\right)=\left\{ s\in S\mid\exists s'\in S'\,:\, s'\overset{\sigma}{\Rightarrow}s\right\}.$
\item All complete initial traces of $\mathcal{A}$ are denoted by $\mathit{ctraces}\left(\mathcal{A}\right)$, which is defined as the set
\[
\left\{ \mathit{trace}\left(\pi\right)\mid\pi\in\mathit{Path\left(\mathcal{A}\right)}:\mathit{first}\left(\pi\right)=s_{0},\left|\pi\right|=\infty\vee\forall a\in L:\mathit{reach}_{\mathcal{A}}\left(\mathit{last}\left(\pi\right),a\right)=\emptyset\right\} .
\]
\item We write $\mathit{after}{}_{\mathcal{A}}\left(s\right)$ for the set
of actions, % that are 
% fix for overfull hbox
enabled from state $s$, i.e. $\mathit{after}{}_{\mathcal{A}}\left(s\right)=\left\{ a\in L\mid s\rightarrow a\right\}$.
We lift this definition to traces by defining 
\[\mathit{after}{}_{\mathcal{A}}\left(\sigma\right)=\bigcup_{s\in\mathit{reach}{}_{\mathcal{A}}\left(s_{0},\sigma\right)}\mathit{after}{}_{\mathcal{A}}\left(s\right).\]

\item We write $\mathit{out}{}_{\mathcal{A}}\left(\sigma\right)=\mathit{after}{}_{\mathcal{A}}\left(\sigma\right)\cap L_{O}^{\delta}$ 
to denote the set of all output actions as well as quiescence after
trace $\sigma$.
\end{itemize}
\end{defn}

In order for a pQTS to be meaningful, \cite{STS13} postulated four well-formedness rules about quiescence, stating for instance that quiescence should not be succeeded by an output action. Since our current treatment does not rely on well-formedness, we omit these rules here. Moreover, our definition of a test case is a pQTS that does not adhere to the well-formedness criteria.

\subsection{Trace distributions}

Very much like the visible behaviour of a labelled transition system is given by its traces, the visible behaviour of a pQTS is given by its trace distributions: each trace distribution is a probability space that assigns a probability to (sets of) traces \cite{Segala:1996:MVR:239648}. Just as a trace in an LTS is obtained by first selecting a path in the LTS and by then removing all states and internal actions, we do the same in the probabilistic case: we first resolve all the nondeterministic choices in the pQTS via 
an adversary, and by then removing all states --- recall that our pQTSs do not contain internal actions. The resolution of the nondeterminism via an adversary leads to a purely probabilistic structure where we can assign a probability to each finite path, by multiplying the probabilities along that path.  The mathematics to handle infinite paths is more complex, but completely standard \cite{cohn1980measure}: in non-trivial situations, the probability assigned to an individual trace is 0 (cf., the probability to always roll a 6 with a dice is 0). Hence, we consider the probability assigned to sets of traces (e.g., the probability that a 6 turns up in the first 100 dice rolls). A classical result in measure theory shows that it is impossible to assign a probability to all sets of traces. Therefore, we collect those sets that can be assigned a probability in a so-called $\sigma$-field $\mathcal{F}$.

\paragraph{Adversaries.}
Following the standard theory for probabilistic automata \cite{MyThesis}, we define the behavior of a pQTS via adversaries (a.k.a. policies or schedulers). These resolve the nondeterministic choices in pQTSs: in each state of the pQTSs, the adversary chooses which transition to take. Adversaries can be (1) history-dependent, i.e. the choice which transition to take can depend on the full history; (2) randomized, i.e. the adversary may make a random choice over all outgoing transitions; and (3) partial, i.e., at any point in time, a scheduler may decide, with some probability, to terminate the execution. 

Thus, given any finite history leading to a current state, an adversary returns a discrete
probability distribution over the set of available next transitions (distributions to be precise). In order to model termination, we define schedulers which continue the transitions of pQTSs with a halting extension. 

\begin{defn}
(Adversary) A \textit{(partial, randomized, history-dependent) adversary}
$E$ of a pQTS $\mathcal{A}=\left(S,s_{0},L_{I},L_{O},\Delta\right)$ is
a function
\[
E:\mathit{Path}{}^{*}\left(\mathcal{A}\right)\longrightarrow\mathit{Distr}\left(\mathit{Distr}\left(L\times S\right)\cup\left\{\perp\right\}\right),
\]
such that for each finite path $\pi$, if $E\left(\pi\right)\left(\mu\right)>0$,
then $\left(last\left(\pi\right),\mu\right)\in\Delta$. The value $E\left(\pi\right)\left(\perp\right)$ is considered as \textit{interruption/halting}. We say
that $E$ is \textit{deterministic}, if $E\left(\pi\right)$ assigns
the Dirac distribution for every distribution after all $\pi\in\mathit{Path}^{*}\left(\mathcal{A}\right)$.
An adversary $E$ halts on a path $\pi$, if it extends $\pi$ to the halting state $\perp$, i.e.
\[
E\left(\pi\right)\left(\perp\right)=1.
\]
We say that an adversary halts after $k\in\mathbb{N}$
steps, if it halts for every path $\pi$ with $\left|\pi\right|\geq k$.
We denote all such adversaries by $\mathit{Adv}\left(\mathcal{A},k\right)$.
Lastly $E$ is finite, if there exists $k\in\mathbb{N}$ such that
$E\in\mathit{Adv}\left(\mathcal{A},k\right)$. 
\end{defn}

\paragraph{The probability space assigned to an adversary.} 
Intuitively an adversary tosses a coin at every step of the computation,
thus resulting in a purely probabilistic (as opposed to nondeterministic)
computation tree. 
\begin{defn}
(Path Probability) Let $E$ be an adversary of $\mathcal{A}$. The
function $Q^{E}:Path^{*}\left(\mathcal{A}\right)\rightarrow\left[0,1\right]$
is called the \textit{path probability function} and it is defined
by induction. We set $Q^{E}\left(s_{0}\right)=1$ and $Q^{E}\left(\pi\mu as\right)=Q^{E}\left(\pi\right)\cdot E\left(\pi\right)\left(\mu\right)\cdot\mu\left(a,s\right)$.
\end{defn}
Loosely speaking, we follow a finite path in the transition system
and multiply every scheduled probability along the way, resolving every nondeterminism
according to the adversary $E$ to get the ultimate path probability. The
path probability function enables us to define a probability space
associated with an adversary, thus giving every path in a pQTS $\mathcal{A}$
an exact probability.
\begin{defn}
(Adversary Probability Space) Let $E$ be an adversary of $\mathcal{A}$.
The \textit{unique probability space associated to} $E$ is the probability
space $\left(\Omega_{E},\mathcal{F}_{E},P_{E}\right)$ given by.
\begin{enumerate}
\item $\Omega_{E}=\mathit{Path}{}^{\infty}\left(\mathcal{A}\right)$
\item $\mathcal{F}_{E}$ is the smallest $\sigma$-field that contains
the set $\left\{ C_{\pi}\mid\pi\in\mathit{Path}{}^{*}\left(\mathcal{A}\right)\right\} $,
where the cone is defined as $C_{\pi}=\left\{ \pi'\in\Omega_{E}\mid\pi\mbox{ is a prefix of }\pi'\right\} $.
\item $P_{E}$ is the unique probability measure on $\mathcal{F}_{E}$ s.
t. $P_{E}\left(C_{\pi}\right)=Q^{E}\left(\pi\right)$, for all $\pi\in\mathit{Path}^{*}\left(\mathcal{A}\right)$.
\end{enumerate}
The set of all adversaries is denoted by $\mathit{adv}\left(\mathcal{A}\right)$
with $\mathit{adv}\left(\mathcal{A},k\right)$ being the set of adversaries
halting after $k\in\mathbb{N}$ respectively. \end{defn}

\paragraph{Trace distributions.}

As we mentioned, a trace distribution is obtained from (the probability space assigned to) an adversary by removing all states. This means that the probability assigned to a set of traces $X$ is defined as the probability of all paths whose trace is an element of $X$. 

\begin{defn}
(Trace Distribution) The \textit{trace distribution} $H$ of an adversary $E$,
denoted $H=\mathit{trd}\left(E\right)$ is the probability space $\left(\Omega_{H},\mathcal{F}_{H},P_{H}\right)$
given by 
\begin{enumerate}
\item $\Omega_{H}=L^*_\mathcal{A}$
\item $\mathcal{F}_{H}$ is the smallest $\sigma$- field containing the
set $\left\{ C_{\beta}\mid\beta\in L^{*}_\mathcal{A}\right\} $,
where the cone is defined as $C_{\beta}=\left\{ \beta'\in\Omega_{E}\mid\beta\mbox{ is a prefix of }\beta'\right\} $
\item $P_{H}$ is the unique probability measure on $\mathcal{F}_{H}$ such
that $P_{H}\left(X\right)=P_{E}\left(\mathit{trace}{}^{-1}\left(X\right)\right)$
for $X\in\mathcal{F}_{H}$.
\end{enumerate}
As an abbreviation, we will write $P_{H}\left(\beta\right):=P_{H}\left(C_{\beta}\right)$ for $\beta\in L^*_{\mathcal{A}}$
\end{defn}
Like before, we denote the set of trace distributions based on adversaries
of $\mathcal{A}$ by $\mathit{trd}\left(\mathcal{A}\right)$ and $\mathit{trd}\left(\mathcal{A},k\right)$
if it is based on an adversary halting after $k\in\mathbb{N}$ steps
respectively. Lastly we write $\mathcal{A}=_{\mathit{TD}}\mathcal{B}$
if $\mathit{trd}\left(\mathcal{A}\right)=\mathit{trd}\left(\mathcal{B}\right)$,
$\mathcal{A}\sqsubseteq_{\mathit{TD}}\mathcal{B}$ if $\mathit{trd}\left(\mathcal{A}\right)\subseteq\mathit{trd}\left(\mathcal{B}\right)$ and $\mathcal{A}\sqsubseteq_{\mathit{TD}}^{k}\mathcal{B}$ if $\mathit{trd}\left(\mathcal{A},k\right)\subseteq\mathit{trd}\left(\mathcal{B},k\right)$
for $k\in\mathbb{N}$, where the embedding means that for every trace distribution $H$ of $\mathcal{A}$ there is a trace distribution $H'$ of $\mathcal{B}$ such that for all traces $\sigma$ of $\mathcal{A}$, we have $P_H\left(\sigma\right)=P_{H'}\left(\sigma\right)$.

The fact that $\left(\Omega_{E},\mathcal{F}_{E},P_{E}\right)$, $\left(\Omega_{H},\mathcal{F}_{H},P_{H}\right)$
really define probability spaces, follows from standard measure theory
arguments (see \cite{cohn1980measure}).

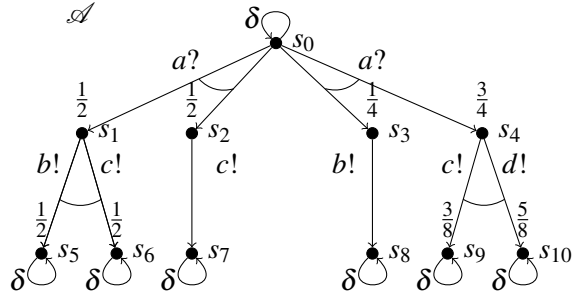
\begin{figure}
\begin{center}
\begin{tikzpicture}[scale=0.8] {}
\node[fill, circle,draw=black, inner sep=1.5, minimum size=0.1, right, label=right:$s_5$, label=above:$\frac{1}{2}$, label=225:$\delta$]  (BottomRowOne) at (0,0)  {};
\node[fill, circle,draw=black, inner sep=1.5, minimum size=0.1, right, label=right:$s_6$, label=above:$\frac{1}{2}$, label=225:$\delta$] (BottomRowTwo) at (1.25,0) {};
\node[fill, circle,draw=black, inner sep=1.5, minimum size=0.1, right, label=right:$s_7$, label=225:$\delta$] (BottomRowThree) at (2.5,0) {};
\node[fill, circle,draw=black, inner sep=1.5, minimum size=0.1, right, label=right:$s_8$, label=225:$\delta$] (BottomRowFour) at (5.5,0) {};
\node[fill, circle,draw=black, inner sep=1.5, minimum size=0.1, right, label=right:$s_9$, label=above:$\frac{3}{8}$, label=225:$\delta$] (BottomRowFive) at (6.75,0) {};
\node[fill, circle,draw=black, inner sep=1.5, minimum size=0.1, right, label=right:$s_{10}$, label=above:$\frac{5}{8}$, label=225:$\delta$] (BottomRowSix) at (8,0) {};

\node[fill, circle,draw=black, inner sep=1.5, minimum size=0.1, right, label=right:$s_1$, label=above:$\frac{1}{2}$] (MiddleRowOne) at (0.675,2) {};
\node[fill, circle,draw=black, inner sep=1.5, minimum size=0.1, right, label=right:$s_2$, label=above:$\frac{1}{2}$] (MiddleRowTwo) at (2.5,2) {};
\node[fill, circle,draw=black, inner sep=1.5, minimum size=0.1, right, label=right:$s_3$, label=above:$\frac{1}{4}$] (MiddleRowThree) at (5.5,2) {};
\node[fill, circle,draw=black, inner sep=1.5, minimum size=0.1, right, label=right:$s_4$, label=above:$\frac{3}{4}$] (MiddleRowFour) at (7.325,2) {};

\node[fill, circle,draw=black, inner sep=1.5, minimum size=0.1, label=right:$s_0$, label=135:$\delta$] (TopRowOne) at (4,3.5) {};

\draw[<-] (BottomRowOne) to (MiddleRowOne);
\draw[<-] (BottomRowTwo) to (MiddleRowOne);

\draw[<-] (BottomRowThree) to (MiddleRowTwo);
\draw[<-] (BottomRowFour) to (MiddleRowThree);

\draw[<-] (BottomRowFive) to (MiddleRowFour);
\draw[<-] (BottomRowSix) to (MiddleRowFour);

\draw[<-] (MiddleRowOne) to (TopRowOne);
\draw[<-] (MiddleRowTwo) to (TopRowOne);

\draw[<-] (MiddleRowThree) to (TopRowOne);
\draw[<-] (MiddleRowFour) to (TopRowOne);

\draw[<-] (BottomRowOne) to (MiddleRowOne);
\draw[<-] (BottomRowTwo) to (MiddleRowOne);

%% Ende des Graphen...dueduedueee %%
%% Arcs #1

\node[left, label=above:$b!$] (MiddlePointOne) at (0.395,1) {};
\node[right, label=above:$c!$] (MiddlePointTwo) at (1.1,1) {};
\draw[-] (MiddlePointOne) to[bend right] (MiddlePointTwo);

%% Arcs #2

\node[right, label=above:$c!$] (MiddlePointThree) at (6.77,1) {};
\node[right, label=above:$d!$] (MiddlePointFour) at (7.8,1) {};
\draw[-] (MiddlePointThree) to[bend right] (MiddlePointFour);

%% Arcs #3
\node[above] (MiddlePointFive) at (2.5,2.9) {$a?$};
\node[right] (MiddlePointSix) at (3.3,2.75) {};
\draw[-] (MiddlePointFive) to[bend right] (MiddlePointSix);

%% Arcs #4
\node[] (MiddlePointFive) at (4.65,2.75) {};
\node[above] (MiddlePointSix) at (5.6,2.9) {$a?$};
\draw[-] (MiddlePointFive) to[bend right] (MiddlePointSix);

%% various labels

\node[right, label=45:$c!$] (Various1) at (2.5,1) {};
\node[right, label=135:$b!$] (Various2) at (5.5,1) {};
\node[right, label=45:$\mathcal{A}$] (Various3) at (0,3.5) {};

%% self loops

\draw[->] (TopRowOne) to [out=45,in=135,looseness=15] (TopRowOne);

\draw[->] (BottomRowOne) to [out=225,in=315,looseness=15] (BottomRowOne);
\draw[->] (BottomRowTwo) to [out=225,in=315,looseness=15] (BottomRowTwo);
\draw[->] (BottomRowThree) to [out=225,in=315,looseness=15] (BottomRowThree);
\draw[->] (BottomRowFour) to [out=225,in=315,looseness=15] (BottomRowFour);
\draw[->] (BottomRowFive) to [out=225,in=315,looseness=15] (BottomRowFive);
\draw[->] (BottomRowSix) to [out=225,in=315,looseness=15] (BottomRowSix);

\end{tikzpicture}
\end{center}
\protect\caption{\label{fig:(ExampleGraph)}An example of the combination of nondeterministic
and probabilistic choices.}
\end{figure}
\begin{example}
Consider the pQTS $\mathcal{A}=\left(S,s_{0}.L_{I},L_{O}^{\delta},\Delta\right)$
in Figure \ref{fig:(ExampleGraph)}. There $S=\left\{ s_{0},s_{1},\ldots,s_{10}\right\} $,
$L_{I}=\left\{ a?\right\} $, $L_{O}^{\delta}=\left\{ b!,c!,d!\right\} \cup\left\{ \delta\right\} $
and $\Delta=\left\{ \left(s_{0},\mu_{0_{1}}\right),\left(s_{0},\mu_{0_{2}}\right),\left(s_{0},\mu_{0_{3}}\right),\left(s_{1},\mu_{1}\right),\ldots,\left(s_{10},\mu_{10}\right)\right\} $.
We can see that this system has both probabilistic and nondeterministic
choices. Observe that it has indeed only input reactive and output
generative transitions as mentioned in the beginning of \ref{subsec:pqts}. 

We will now consider an adversary $E$ for $\mathcal{A}$. The only
nondeterministic choice we have in this system, is located at state
$s_{0}$, where we can either apply $a?$ to enter the left branch,
$a?$ to enter the right branch, or do nothing (corresponding to $\mu_{0_{1}}$,
$\mu_{0_{2}}$ and $\mu_{0_{3}}$ respectively). Therefore consider
the adversary $E\left(s_{0}\right)\left(\mu_{0_{1}}\right)=\frac{1}{2}$
and $E\left(s_{0}\right)\left(\mu_{0_{2}}\right)=\frac{1}{2}$ and
$E\left(\pi\right)\left(\mu\right)=\mathit{Dirac}$ for every other
distribution $\mu$ after a path $\pi$ (i.e. those are taken with
probability $1$).

The adversary probability space created for this adversary assigns
an unambiguous path probability to each path. Consider the path $\pi=s_{0}\mu_{0_{1}}a?s_{1}\mu_{1}b!s_{5}$,
then 
\[
P_{E}\left(\pi\right)=Q^{E}\left(\pi\right)=\underset{1}{\underbrace{Q^{E}\left(s_{0}\right)}}\underset{\frac{1}{2}}{\underbrace{E\left(s_{0}\right)\left(\mu_{0_{1}}\right)}}\underset{\frac{1}{2}}{\underbrace{\mu_{0_{1}}\left(a?,s_{1}\right)}}\underset{1}{\underbrace{E\left(s_{0}\mu_{0_{1}}a?s_{1}\right)\left(\mu_{1}\right)}}\underset{\frac{1}{2}}{\underbrace{\mu_{1}\left(b!,s_{5}\right)}}=\frac{1}{8}.
\]
However, consider the trace distribution $H=\mathit{trd}\left(E\right)$.
Then for $\sigma=a?b!$, we have $\mathit{trace}^{-1}\left(\sigma\right)=\left\{ \pi,\eta\right\} $
with $\pi$ as before and $\eta=s_{0}\mu_{0_{2}}a?s_{3}\mu_{3}b!s_{8}$.
Hence 
\begin{eqnarray*}
P_{H}\left(\sigma\right) & = & P_{\mathit{trd}\left(E\right)}\left(\mathit{trace}^{-1}\left(\sigma\right)\right)=P_{E}\left(\left\{ \pi,\eta\right\} \right)=P_{E}\left(\pi\right)+P_{E}\left(\eta\right)=\frac{1}{4}.
\end{eqnarray*}

\end{example}

\section{The probabilistic conformance relation \pioco}\label{sec(pioco)}

\subsection{The \pioco relation}

The classical input-output conformance relation \ioco states that an implementation $\mathcal{A}_{i}$ conforms to a specification $\mathcal{A}_{s}$  
if $\mathcal{A}_{i}$ never provides any unspecified output. In particular this refers to the observation of quiescence, when other output was expected. 

\begin{defn}
(Input- Output Conformance) Let $\mathcal{A}_{i}$ and $\mathcal{A}_{s}$
be two QTS and let $\mathcal{A}_{i}$ be input enabled. Then we say
$\mathcal{A}_{i}\sqsubseteq_{ioco}\mathcal{A}_{s}$, if and only if
\[
\forall\sigma\in\mathit{traces}\left(\mathcal{A}_{s}\right):\mathit{out}_{\mathcal{A}_{i}}\left(\sigma\right)\subseteq\mathit{out}_{\mathcal{A}_{s}}\left(\sigma\right).
\]

\end{defn}

To generalize \ioco to pQTSs, we introduce two auxiliary concepts. For a natural number $k$, the prefix relation $H \sqsubseteq_{k} H'$ states that trace distribution $H$ assigns exactly the same probabilities as $H'$ to traces of length $k$ and halts afterwards.
The output continuation of a trace distribution $H$ prolongs the traces of $H$ with output actions. More precisely, output continuation of $H$ wrt length $k$ contains all trace distributions that (1) coincide with $H$ for traces upto length $k$ and (2) the $k+1$st action is an output label (incl $\delta$); i.e. traces of length $k+1$ that end on an input action are assigned probability 0. Recall that $P_{H}\left(\sigma\right)$ abbreviates $P_{H}\left(C_\sigma\right)$. 

\begin{defn}
(Notations) For a natural number $k\in\mathbb{N}$, 
and trace distributions $H \in\mathit{trd}\left(\mathcal{A},k\right)$, we say that 
\begin{enumerate}
\item $H$
is a {\em prefix} of $H'\in\mathit{trd}\left(\mathcal{A}\right)$ up to $k$, denoted by $H\sqsubseteq_{k}H'$, iff
$\forall\sigma\in L^{k}:P_{H}\left(\sigma\right)=P_{H'}\left(\sigma\right).$ 

\item the {\em output continuation} of $H$ in $\mathcal{A}$ is given by
\begin{eqnarray*}
\mathit{outcont}\left(H,\mathcal{A},k\right): & = & \left\{ H'\in\mathit{trd}\left(\mathcal{A},k+1\right) \mid H\sqsubseteq_{k}H'\wedge\forall\sigma\in L^{k}L_{I}:P_{H'}\left(\sigma\right)=0\right\} .
\end{eqnarray*}
\end{enumerate}
\end{defn}
We are now able to define the core idea of \pioco. Intuitively an implementation should conform to a specification, if the probability of every trace in $\mathcal{A}_i$ specified in $\mathcal{A}_s$, can be matched in the specification. Just as in \ioco , we will neglect underspecified traces continued with input actions (i.e., everything is allowed to happen after that). However, if there is unspecified output in the implementation, there is at least one adversary that schedules positive probability to this continuation, which consequently cannot be matched of output-continuations in the specification.
\begin{defn}
Let $\mathcal{A}_{i}$ and $\mathcal{A}_{s}$ be two pQTS. Furthermore let $\mathcal{A}_i$ be input enabled, then we
say $\mathcal{A}_{i}\sqsubseteq_{\mathit{pioco}}\mathcal{A}_{s}$
if and only if

\[
\forall k\in\mathbb{N}\forall H\in\mathit{trd}\left(\mathcal{A}_{s},k\right):\mathit{outcont}\left(H,\mathcal{A}_{i},k\right)\subseteq\mathit{outcont}\left(H,\mathcal{A}_{s},k\right).
\]
\end{defn}

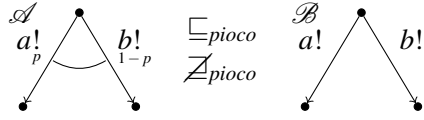
\begin{figure}
\begin{center}\begin{tikzpicture}
\node[fill, circle,draw=black, inner sep=1, minimum size=0.1] (initImp) at (1,1.25) {} ;
\node[fill, circle,draw=black, inner sep=1, minimum size=0.1] (leftImp) at (0.25,0) {};
\node[fill, circle,draw=black, inner sep=1, minimum size=0.1] (rightImp) at (1.75,0) {};

\draw[->] (initImp) to node[inner sep=0,label=135:$a!$, label=left:\tiny $p$](middle1){} (leftImp);
\draw[->] (initImp) to node[inner sep=0,label=45:$b!$, label=right:\tiny $1-p$](middle2){} (rightImp);
\draw (middle1) to[bend right] (middle2);

\node (pioco1) at (2.9,1) {$\sqsubseteq_{\mathit{pioco}}$} ;
\node (pioco2) at (2.9,0.5) {$\cancel{\sqsupseteq}_{\mathit{pioco}}$};

\node (labelA) at (0.25,1.25) {$\mathcal{A}$} ;
\node (labelB) at (4,1.25) {$\mathcal{B}$} ;

\node[fill, circle,draw=black, inner sep=1, minimum size=0.1] (initSpec) at (4.75,1.25) {} ;
\node[fill, circle,draw=black, inner sep=1, minimum size=0.1] (leftSpec) at (4,0) {};
\node[fill, circle,draw=black, inner sep=1, minimum size=0.1] (rightSpec) at (5.5,0) {};

\draw[->] (initSpec) to node[label=135:$a!$](middle3){} (leftSpec);
\draw[->] (initSpec) to node[label=45:$b!$](middle4){} (rightSpec);
\end{tikzpicture}

\end{center}\caption{An example illustrating \pioco}\label{fig:(piocoExample)}\end{figure}
\begin{example} Consider the two systems of $\mathcal{A}$ and $\mathcal{B}$ shown in Figure \ref{fig:(piocoExample)} and assume that $p\in\left[0,1\right]$. It is true that $\mathcal{A}\sqsubseteq_\mathit{\mathit{pioco}}\mathcal{B}$, because we can always choose an adversary $E$ of $\mathcal{B}$, which imitates the probabilistic behaviour of $\mathcal{B}$, i.e. choose $E(\varepsilon)(\mu)=\nu$ such that $\nu\left(a!,t_1\right)=p$ and $\nu\left(b!,t_2\right)=1-p$. 

However, the opposite does not  hold. For example assume $p=\frac{1}{2}$, then the trace distribution $H$ assigning $P_H\left(a!\right)=1$ is in $\mathit{outcont}\left(H,\mathcal{B},1\right)$ but not in $\mathit{outcont}\left(H,\mathcal{A},1\right)$ and hence $\mathcal{B}\cancel{\sqsubseteq}_\mathit{\mathit{pioco}}\mathcal{A}$.

\end{example}

\subsection{Properties of the p-ioco relation}
As stated before, the relation \pioco conservatively extends the \ioco relation, i.e. both relations coincide for non-probabilistic QTSs. Moreover, we show that several other characteristic properties of \ioco carry over to \pioco as well. Below, a QTS is a pQTS where every occurring distribution is the Dirac distribution.

\begin{thm}
Let $\mathcal{A}_{i}$ and $\mathcal{A}_{s}$ be two QTS and let $\mathcal{A}_i$ be input enabled, then
\[
\mathcal{A}_{i}\sqsubseteq_{\mathit{ioco}}\mathcal{A}_{s}\Longleftrightarrow\mathcal{A}_{i}\sqsubseteq_{\mathit{pioco}}\mathcal{A}_{s}.
\]
\end{thm}
\proofatend
\qquad\\%fight overfull -- start the proof on the next line
\indent $"\Longleftarrow"$ \quad Let $\mathcal{A}_{i}\sqsubseteq_{\mathit{pioco}}\mathcal{A}_{s}$
and $\sigma\in\mathit{traces}\left(\mathcal{A}_{s}\right)$. Our goal
is to show $\mathit{out}_{\mathcal{A}_{i}}\left(\sigma\right)\subseteq\mathit{out}_{\mathcal{A}_{s}}\left(\sigma\right)$. 

For $\mathit{out}_{\mathcal{A}_{i}}\left(\sigma\right)=\emptyset$
we are done, since $\emptyset\subseteq\mathit{out}_{\mathcal{A}_{s}}\left(\sigma\right)$
obviously. 

So assume that there is $b!\in\mathit{out}_{\mathcal{A}_{i}}\left(\sigma\right)$.
We want to show that $b!\in\mathit{out}_{\mathcal{A}_{s}}\left(\sigma\right)$.
For this, let $k=\left|\sigma\right|$ and $H\in\mathit{trd}\left(\mathcal{A}_{s},k\right)$
such that $P_{H}\left(\sigma\right)=1$, which is possible because
$\sigma\in\mathit{traces}\left(\mathcal{A}_{s}\right)$ and both $\mathcal{A}_{i}$
and $\mathcal{A}_{s}$ are non-probabilistic. The same argument gives
us $\mathit{outcont}\left(H,\mathcal{A}_{i},k\right)\neq\emptyset$,
because $\sigma\in\mathit{traces}\left(\mathcal{A}_{i}\right)$. 

Thus we have at least one $H'\in\mathit{outcont}\left(H,\mathcal{A}_{i},k\right)$
such that $P_{H'}\left(\sigma b!\right)>0$. Let $\pi\in\mathit{trace}^{-1}\left(\sigma\right)\cap\mathit{Path^{*}}\left(\mathcal{A}_{s}\right)$.
Now $H'\in\mathit{outcont}\left(H,\mathcal{A}_{s},k\right)$, because
$\mathcal{A}_{i}\sqsubseteq_{\mathit{pioco}}\mathcal{A}_{s}$ by assumption
and thus there must be at least one adversary $E'\in\mathit{adv}\left(\mathcal{A}_{s},k+1\right)$
such that $\mathit{trd}\left(E'\right)=H'$ and $Q^{E'}\left(\pi\cdot\mathit{Dirac}\cdot b!s'\right)>0$
for some $s'\in S$. Hence $E'\left(\pi\right)\left(\mathit{Dirac}\right)\mathit{Dirac}\left(b!,s'\right)>0$
and therefore with $s'\in\mathit{reach}\left(\mathit{last}\left(\pi\right),b!\right)$
this yields $b!\in\mathit{out}_{\mathcal{A}_{s}}\left(\sigma\right)$.\\

$"\Longrightarrow"$ Let $\mathcal{A}_{i}\sqsubseteq_{ioco}\mathcal{A}_{s}$,
$k\in\mathbb{N}$ and $H^{*}\in\mathit{trd}\left(\mathcal{A}_{s},k\right)$.
Assume that $H\in\mathit{outcont}\left(H^{*},\mathcal{A}_{i},k\right)$,
then we want to show that $H\in\mathit{outcont}\left(H^{*},\mathcal{A}_{s},k\right)$. 

Therefore let $E\in\mathit{adv}\left(\mathcal{A}_{i},k+1\right)$
such that $\mathit{trd}\left(E\right)=H$. If we can find $E'\in\mathit{adv}\left(\mathcal{A}_{s},k+1\right)$
such that $\mathit{trd}\left(E\right)=\mathit{trd}\left(E'\right)$,
we are done. We will do this constructively in three steps.\\
\\
1) By construction of $H^{*}$ we know that there must be $E'\in\mathit{adv}\left(\mathcal{A}_{s},k+1\right)$,
such that for all $\sigma\in L^{k}$ we have $P_{\mathit{trd}\left(E'\right)}\left(\sigma\right)=P_{H^{*}}\left(\sigma\right)=P_{\mathit{trd}\left(E\right)}\left(\sigma\right)$.
Thus $H^{*}\sqsubseteq_{k}\mathit{trd}\left(E'\right)$.\\
\\
2) We did not specify the behaviour of $E'$ for path of length $k+1$.
Therefore we choose $E'$ such that for all traces $\sigma\in L^{k}$
and $a?\in L_{I}$ we have $P_{\mathit{trd}\left(E'\right)}\left(\sigma a?\right)=0=P_{\mathit{trd}\left(E\right)}\left(\sigma a?\right)$.
\\
\\
3) The last thing to show is that $\mathit{trd}\left(E\right)=\mathit{trd}\left(E'\right)$.
Therefore let us now set the behaviour of $E'$ for traces ending
in outputs. Let $\sigma\in\mathit{traces}\left(\mathcal{A}_{i}\right)$,
then assume $a!\in\mathit{out}_{\mathcal{A}_{i}}\left(\sigma\right)$
(if $\mathit{out}_{\mathcal{A}_{i}}\left(\sigma\right)=\emptyset$
we are done immediately) and because $\mathcal{A}_{i}\sqsubseteq_{\mathit{ioco}}\mathcal{A}_{s}$,
we know that $a!\in\mathit{out}_{\mathcal{A}_{s}}\left(\sigma\right)$. 

Now let $p:=P_{\mathit{trd}\left(E\right)}\left(\sigma\right)=P_{\mathit{trd}\left(E'\right)}\left(\sigma\right)$
and $q:=P_{\mathit{trd}\left(E\right)}\left(\sigma a!\right)$. By
equality of the trace distributions for traces up to length $k$ we
know that $q\leq p\leq1$ and therefore there is $\alpha\in\left[0,1\right]$
such that $q=p\cdot\alpha$. Let $\mathit{traces}\left(\mathcal{A}_{s}\right)\cap\mathit{trace}^{-1}\left(\sigma\right)=\left\{ \pi_{1},\ldots,\pi_{n}\right\} $.
Without loss of generality, we choose $E'$ such that 
\[
E'\left(\pi_{i}\right)\left(\mathit{Dirac}\right)=\begin{cases}
\alpha & \mbox{ if }i=1\\
0 & \mbox{ else}
\end{cases}.
\]% %fight overfull -- shorten beginnign of the sentence
%In this way we 
We constructed $E'\in\mathit{adv}\left(\mathcal{A}_{s},k+1\right)$,
such that for all $\sigma\in L^{k+1}$ we have $P_{\mathit{trd}\left(E'\right)}\left(\sigma\right)=P_{\mathit{trd}\left(E\right)}\left(\sigma\right)$
and thus $\mathit{trd}\left(E\right)=\mathit{trd}\left(E'\right)$,
which finally yields $H\in\mathit{outcont}\left(H^{*},\mathcal{A}_{s},k\right)$.
\endproofatend
Intuitively it makes sense that the implementation is input enabled, since it should accept every input at any time. The following two results justify, that we assume the specification to be not input enabled, since otherwise \pioco would coincide with trace distribution inclusion. Equivalently it is known that \ioco coincides with trace inclusion, if we assume both the implementation and the specification were input enabled. Thus, as stated before, we can see that \pioco extends \ioco.

\begin{lem}
\label{lem:(weak implication)}Let $\mathcal{A}_{i}$ and $\mathcal{A}_{s}$
be two pQTS, then

\[
\mathcal{A}_{i}\sqsubseteq_{\mathit{TD}}\mathcal{A}_{s}\Longrightarrow\mathcal{A}_{i}\sqsubseteq_{\mathit{pioco}}\mathcal{A}_{s}.
\]
\end{lem}
\proofatend
Let $\mathcal{A}_{i}\sqsubseteq_{\mathit{TD}}^{k}\mathcal{A}_{s}$
then for every $H\in\mathit{trd}\left(\mathcal{A}_{i},k\right)$ we
also have $H\in\mathit{trd}\left(\mathcal{A}_{s},k\right)$. So pick
$m\in\mathbb{N}$, let $H^{*}\in\mathit{trd}\left(\mathcal{A}_{s},m\right)$
and take $H\in\mathit{outcont}\left(H^{*},\mathcal{A}_{i},m\right)\subseteq\mathit{trd}\left(\mathcal{A}_{i},m+1\right)$.
We want to show that $H\in\mathit{outcont}\left(H^{*},\mathcal{A}_{s},m\right)$. 

By assumption we know that $H\in\mathit{trd}\left(\mathcal{A}_{s},m+1\right)$.
In particular that means there must be at least one adversary $E\in\mathit{adv}\left(\mathcal{A}_{s},m+1\right)$
such that $\mathit{trd}\left(E\right)=H$. However, for this adversary,
we know that $H^{*}\sqsubseteq_{m}\mathit{trd}\left(E\right)$ and
for all $\sigma\in L^{m}L_{I}$ we have $P_{\mathit{trd}\left(E\right)}\left(\sigma\right)=0$
and by trace distribution inclusion $\mathit{trd}\left(E\right)=H$.
Thus $H\in\mathit{outcont}\left(H^{*},\mathcal{A}_{s},m\right)$ and
therefore $\mathcal{A}_{i}\sqsubseteq_{\mathit{pioco}}\mathcal{A}_{s}$.\endproofatend
\begin{thm}
\label{thm:(pioco =000026 TD)}Let $\mathcal{A}_{i}$ and $\mathcal{A}_{s}$
be two input enabled pQTS, then 

\[
\mathcal{A}_{i}\sqsubseteq_{\mathit{pioco}}\mathcal{A}_{s}\Longleftrightarrow \mathcal{A}_{i}\sqsubseteq_{TD}\mathcal{A}_{s}.
\]
\end{thm}
\proofatend
$"\Longrightarrow"$ Let $\mathcal{A}_{i}\sqsubseteq_{\mathit{pioco}}\mathcal{A}_{s}$,
fix $m\in\mathbb{N}$ and take a trace distribution $H^{*}\in\mathit{trd}\left(\mathcal{A}_{i},m\right)$.
To show that $H^{*}\in\mathit{trd}\left(\mathcal{A}_{s},m\right)$,
we prove that every prefix of $H^{*}$ is in $\mathit{trd}\left(\mathcal{A}_{s},m\right)$,
i.e. if $H'\sqsubseteq_{k}H^{*}$ for some $k\in\mathbb{N}$, then
$H'\in\mathit{trd}\left(\mathcal{A}_{s}\right)$. The proof is by
induction up to $m\in\mathbb{N}$ over the prefix trace distribution
length, denoted by $k$. 

Obviously $H'\in\mathit{trd}\left(\mathcal{A}_{i},0\right)$ yields
both $H'\sqsubseteq_{0}H^{*}$ and $H'\in\mathit{trd}\left(\mathcal{A}_{s}\right)$.
Now assume, we know that $H'\sqsubseteq_{k}H^{*}$ for some $k<m$
and $H'\in\mathit{trd}\left(\mathcal{A}_{s}\right)$. Furthermore
let $H''\in\mathit{trd}\left(\mathcal{A}_{i},k+1\right)$, such that
$H''\sqsubseteq_{k+1}H^{*}$. If we can show that $H''\in\mathit{trd}\left(\mathcal{A}_{s},k+1\right)$,
we are done. 

With $H'\in\mathit{trd}\left(\mathcal{A}_{s},k\right)$, we take $H'''\in\mathit{outcont}\left(H',\mathcal{A}_{i},k\right)$
such that all traces of length $k+1$ ending in an output action have
the same probability, i.e. for all $\sigma\in L^{k}L_{O}^{\delta}$,
we have $P_{H'''}\left(\sigma\right)=P_{H''}\left(\sigma\right)$.
By assumption $\mathcal{A}_{i}\sqsubseteq_{\mathit{pioco}}\mathcal{A}_{s}$
and thus $H'''\in\mathit{outcont}\left(H',\mathcal{A}_{s},k\right)\subseteq\mathit{trd}\left(\mathcal{A}_{s}\right)$. 

Let $E\in\mathit{adv}\left(\mathcal{A}_{s},k+1\right)$ the corresponding
adversary such that $\mathit{trd}\left(E\right)=H'''$. By construction,
we have $P_{\mathit{trd}\left(E\right)}\left(\sigma a!\right)=P_{H''}\left(\sigma a!\right)$
and $P_{\mathit{trd}\left(E\right)}\left(\sigma b?\right)=0\overset{\tiny\mbox{in general}}{\neq}P_{H''}\left(\sigma b?\right)$
for all $\sigma\in L^{k}$. We create yet another adversary, denoted
by $E'\in\mathit{adv}\left(\mathcal{A}_{s},k+1\right)$ such that
for all $\sigma\in L^{k}$ and $a!\in L_{O}^{\delta}$, we have $P_{\mathit{trd}\left(E\right)}\left(\sigma\right)=P_{\mathit{trd}\left(E'\right)}\left(\sigma\right)$
and $P_{\mathit{trd}\left(E\right)}\left(\sigma a!\right)=P_{\mathit{trd}\left(E'\right)}\left(\sigma a!\right)$.
Taking the sum over all probabilities of those traces yields 
\[
\sum_{a!\in L_{O}^{\delta}}P_{\mathit{trd}\left(E\right)}\left(\sigma a!\right)=1-\alpha,
\]
where $\alpha\in\left[0,1\right]$ and consequently the remaining
bit is covered by 
\[
\sum_{b?\in L_{I}}P_{H''}\left(\sigma b?\right)=\alpha.
\]
\\
The aim is now to set the behaviour of $E'$ such that $\sigma\in L^{k}L_{I}$
has $P_{H''}\left(\sigma\right)=P_{\mathit{trd}\left(E'\right)}\left(\sigma\right)$.
We prove that this can indeed be done independently from $\sigma$.
The input enabledness gives that for all $\sigma b?\in\mathit{traces}\left(\mathcal{A}_{i}\right)$,
we also have $\sigma b?\in\mathit{traces}\left(\mathcal{A}_{s}\right)$.
Assume $P_{H''}\left(\sigma\right)=p$ and thus 
\begin{eqnarray*}
\alpha & = & \sum_{b?\in L_{I}}P_{H''}\left(\sigma b?\right)=P_{H''}\left(\sigma b_{1}?\right)+\ldots+P_{H''}\left(\sigma b_{n}?\right)=p\alpha_{1}+\ldots+p\alpha_{\omega}\\
 & \overset{!}{=} & P_{\mathit{trd}\left(E'\right)}\left(\sigma b_{1}?\right)+\ldots+P_{\mathit{trd}\left(E'\right)}\left(\sigma b_{n}?\right).
\end{eqnarray*}
However, since $\mathit{trd}\left(E\right)\sqsubseteq_{k}H''$, we
also have $P_{\mathit{trd}\left(E\right)}\left(\sigma\right)=p$. 

The last detail not yet specified about $E'$ is the behaviour of
paths of length $k+1$ ending in an input transition. We demonstrate
the choice of $E'$ for $p\alpha_{1}\overset{!}{=}P_{\mathit{trd}\left(E'\right)}\left(\sigma b_{1}?\right)$,
and denote the associated paths $\left\{ \pi_{1},\ldots,\pi_{n}\right\} =\mathit{trace}^{-1}\left(\sigma\right)$.
Furthermore $\pi_{i}':=\pi_{i}\mu b_{1}?s_{i_{j}}$ for some $s_{i_{j}}\in S$,
$j=1,\ldots,l$, which are reachable after $\pi_{i}$ and distributions
containing $b?$. Thus we want

\begin{eqnarray*}
p\alpha_{1} & \overset{!}{=} & P_{\mathit{trd}\left(E'\right)}\left(\sigma b?\right)=\sum_{i=1}^{n}P_{E'}\left(\pi_{i}'\right)\\
 & = & \sum_{i=1}^{n}\sum_{j=1}^{l}\underset{=p}{\underbrace{Q^{E'}\left(\pi_{i}\right)}}\underset{=:\alpha_{1}}{\underbrace{E'\left(\pi_{i}'\right)\left(\mu\right)}}\mu\left(b_{1}?,s_{i_{j}}\right)\\
 & = & p\alpha_{1}\underset{=1}{\underbrace{\sum_{i=1}^{n}\sum_{j=1}^{l}\mu\left(b_{1}?,s_{i_{j}}\right)}.}
\end{eqnarray*}
We can do the same for all $\alpha_{i}$ for $i=1,\ldots,\omega$.
Note that the choice of the adversary does \textit{not} depend on
the chosen trace $\sigma$ but solely on the presupposed behaviour
of $H''$. Thus we have found $E'\in\mathit{adv}\left(\mathcal{A}_{s},k+1\right)$
such that $\mathit{trd}\left(E'\right)=H''$. Hence $H''\in\mathit{trd}\left(\mathcal{A}_{s},k+1\right)$, which ends the induction. Since this is possible for every $m\in\mathbb{N}$, we get $\mathcal{A}_i\subseteq_{\mathit{pioco}}\mathcal{A}_s$, ending the proof.\\

$"\Longleftarrow"$ See Lemma \ref{lem:(weak implication)} for the
proof. In particular we do not even require input enabledness for
$\mathcal{A}_{s}$ in this case.
\endproofatend 
Next, we show that, under some input-enabledness restrictions, the \pioco relation is transitive. Again,  note that 
this is also true for \ioco for non-probabilistic systems.
\begin{thm}
(Transitivity of pioco) Let $\mathcal{A}$, $\mathcal{B}$ and $\mathcal{C}$
be pQTS, such that $\mathcal{A}$ and $\mathcal{B}$ are input enabled,
then 

\[
\mathcal{A}\sqsubseteq_{\mathit{pioco}}\mathcal{B}\wedge\mathcal{B}\sqsubseteq_{\mathit{pioco}}\mathcal{C}\Longrightarrow\mathcal{A}\sqsubseteq_{\mathit{pioco}}\mathcal{C}.
\]
\end{thm}
\proofatend
Let $\mathcal{A}\sqsubseteq_{\mathit{pioco}}\mathcal{B}$ and $\mathcal{B}\sqsubseteq_{\mathit{pioco}}\mathcal{C}$
and $\mathcal{A}$ and $\mathcal{B}$ be input enabled. By Theorem
\ref{thm:(pioco =000026 TD)} we know, that $\mathcal{A}\sqsubseteq_{\mathit{TD}}\mathcal{B}$.
So let $k\in\mathbb{N}$ and $H^{*}\in\mathit{trd}\left(\mathcal{A},k\right)$.
Consequently also $H^{*}\in\mathit{trd}\left(\mathcal{B},k\right)$
and thus the following embedding holds

\[
\mathit{outcont}\left(\mathcal{A},H^{*},k\right)\subseteq\mathit{outcont}\left(\mathcal{B},H^{*},k\right)\subseteq\mathit{outcont}\left(\mathcal{C},H^{*},k\right),
\]
 and thus $\mathcal{A}\sqsubseteq_{\mathit{pioco}}\mathcal{C}$.
\endproofatend

\section{Testing for pQTS}\label{sec:Testing-for-pQTS}

\subsection{Test cases for pQTSs.} We will consider tests as sets of traces based on an action signature $\left(L_I,L_O^\delta\right)$, which will describe possible behaviour of the tester. This means that at each state in a test case, the tester either provides stimuli or waits for a response of the system. Additionally to output conformance testing like in ~\cite{TimmerStoelingaBrinksma}, we introduce probabilities into our testing transition system. Thus we can represent each test case as a pQTS, albeit with a mirrored action signature $\left(L_O,L_I\cup\left\{\delta\right\} \right)$. This is necessary for the parallel composition of the test pQTS and the SUT. 

Since we consider tests to be pQTS, we also use all the terminology introduced earlier on. Additionally we require tests to not contain loops (or infinite paths respectively).	

\begin{defn}
\label{def:(Test)} A \textit{test (directed acyclic graph)}
over an action signature $\left(L_{I},L_{O}^{\delta}\right)$ is a
pQTS of the form $t=\left(S,s_{0},L_{O},L_{I}\cup\left\{ \delta\right\} ,\Delta\right)$
 such that \end{defn}
\begin{itemize}
\item $t$ is internally deterministic and does not contain an infinite
path; 
\item $t$ is acyclic and connected;
\item For every state $s\in S$, we either have\\
- $\mathit{after}\left(s\right)=\emptyset$, or\\
- $\mathit{after}\left(s\right)=L_{I}\cup\left\{ \delta\right\} $,
or \\
- $\mathit{after}\left(s\right)=\left\{ a!\right\} \cup L_{I}\cup\left\{ \delta\right\} $
for some $a!\in L_{O}$.
\end{itemize}
A test suite $T$ is a set of tests over an action signature $\left(L_{I},L_{O}^{\delta}\right)$.
We write $\mathcal{T}\left(L_{I},L_{O}^{\delta}\right)$ to denote
all the tests over an action signature $\left(L_{I},L_{O}^{\delta}\right)$
and $\mathcal{TS}\left(L_{I},L_{O}^{\delta}\right)$ as the set of
all test suites over an action signature respectively. 

For a given specification pQTS $\mathcal{A}_{s}=\left(S,s_{0},L_{I},L_{O}^{\delta},\Delta\right)$,
we say that a test $t$ is a \textit{test for} $\mathcal{A}_{s}$,
if it is based on the same action signature $\left(L_{I},L_{O}^{\delta}\right)$.
Similar to before, we denote all tests for $\mathcal{A}_{s}$ by $\mathcal{T}\left(\mathcal{A}_{s}\right)$
and all test suites by $\mathcal{TS}\left(\mathcal{A}_{s}\right)$
respectively.

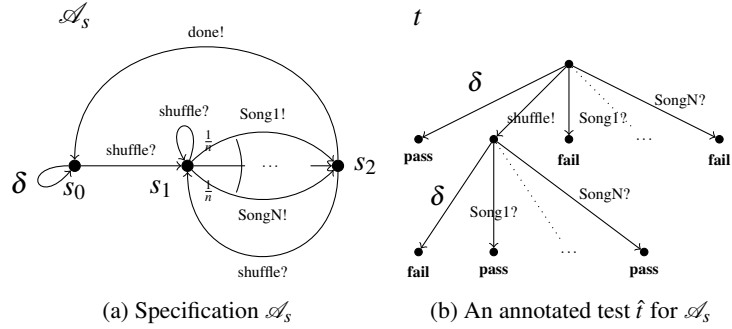
\begin{figure}
\begin{center}\subfloat[\label{fig:Specification}Specification $\mathcal{A}_{s}$]{\begin{tikzpicture}

\node[fill, circle,draw=black, inner sep=1.5, minimum size=0.1, label=below:$s_0$] (init) at (0,0) {};
\node[fill, circle,draw=black, inner sep=1.5, minimum size=0.1, label=225:$s_1$, label=45:\tiny$\frac{1}{n}$, label=315:\tiny$\frac{1}{n}$] (select) at (1.5,0) {};
\node[fill, circle,draw=black, inner sep=1.5, minimum size=0.1, label=right:$s_2$] (done) at (3.5,0) {};

\node (arcBegin) at (2.1,0.49) {};
\node (arcEnd) at (2.1,-0.49) {};

\node (label) at (0,2) {$\mathcal{A}_s$};

\node (afterSelect) at (2.4,0) {};
\node (beforeDone) at (3.0,0) {};
\node (ldots) at (2.6,0) {\tiny$\ldots$};

%% Drawing the transitions
\draw[-] (select) to (afterSelect);
\draw[->] (beforeDone) to (done);

\draw[->] (init) to node[above]{\tiny shuffle?} (select);
\draw[->] (select) to[in=-225, out=45] node[above]{\tiny  Song1!} (done);
\draw[->] (select) to[in=225,  out=-45] node[below]{\tiny SongN!}(done);
\draw[->] (done) to[in=-90,  out=-90, distance=1.5cm] node[below]{\tiny shuffle?}(select);
\draw[->] (done) to[in=90,  out=90, distance=2cm] node[above]{\tiny done!}(init);
\draw[-] (arcBegin) to[in=70, out=290] (arcEnd);

\draw[->] (select) to [in=120, out=70, looseness=25] node[above] {\tiny shuffle?} (select);
\draw[->] (init) to [out=180, in=230, looseness=25] node[left] {$\delta$} (init);
\end{tikzpicture}

}\subfloat[\label{fig:Test}An annotated test $\hat{t}$ for $\mathcal{A}_{s}$]{\begin{tikzpicture}
\tikzset{standart/.style={fill, circle,draw=black, inner sep=1, minimum size=0.1}};
\node[standart] (init) at (0,0) {};
\node[standart, label=below:\tiny$\textbf{pass}$] (firstFirst) at (-2,-1) {};
\node[standart] (firstSecond) at (-1,-1) {};
\node[standart, label=below:\tiny$\textbf{fail}$] (firstThird) at (0,-1) {};
%\node[label=center:\tiny $\ldots$] (firstFourth) at (1,-1) {};
\node[] (firstFourth) at (1,-1) {\tiny $\ldots$};
\node[standart, label=below:\tiny$\textbf{fail}$] (firstFifth) at (2,-1) {};
\node[label=below:$t$] (label) at (-2,1) {};
\node[standart, label=below:\tiny$\textbf{fail}$] (secondFirst) at (-2,-2.5) {};
\node[standart, label=below:\tiny$\textbf{pass}$] (secondSecond) at (-1,-2.5) {};
%\node[label=center:\tiny $\ldots$] (secondThird) at (0,-2.5) {};
\node[] (secondThird) at (0,-2.5) {\tiny $\ldots$};
\node[standart, label=below:\tiny$\textbf{pass}$] (secondFourth) at (1,-2.5) {};
%%draw first row
\draw[->] (init) to node[above left] {\small $\delta$} (firstFirst);
\draw[->] (init) to node[below] {\tiny shuffle!} (firstSecond);
\draw[->] (init) to node[below right] {\tiny Song1?} (firstThird);
\draw[dotted] (init) to (firstFourth);
\draw[->] (init) to node[right] {\tiny SongN?} (firstFifth);

%%%%%%%%%%%%%%%%%%%%%%%%%%%
%draw second row
\draw[->] (firstSecond) to node[left] {\small $\delta$} (secondFirst);
\draw[->] (firstSecond) to node[below] {\tiny Song1?} (secondSecond);
\draw[dotted] (firstSecond) to  (secondThird);
\draw[->] (firstSecond)  to node[right] {\tiny SongN?} (secondFourth);

%%\node
\end{tikzpicture}

}\end{center}\protect\caption{\label{fig:SpecAndTest}A specification for a simple shuffle music
player and a test.}

\end{figure}

Note that we mirrored the action signature for tests, as can be seen
in Figure \ref{fig:Specification} and Figure \ref{fig:Test} respectively.
That is, because we require tests and implementations to shake hands
on shared actions. A special role is dedicated to quiescence in the
context of parallel composition, since the composed system is considered
quiescent if and only if the two systems are quiescent. 

We will proceed to define parallel composition. Formally this means that output actions of one component are allowed to be present as input actions of the other component. These will be synchronized upon. However, keeping in mind the mirrored action signature of tests, we wish to avoid possibly unwanted synchronization, which is why we introduce system compatibility.  

\begin{defn}
(Compatibility) Two pQTS 
$
\mathcal{A}=\left( S,s_{0},L_{I},L_{O}^{\delta},\Delta\right) ,
$ and 
$
\mathcal{A}'=\left( S',s_{0}',L_{I}',L_{O}^{\delta\prime},\Delta'\right) 
$
 are said to be \textit{compatible} if $L_{O}^{\delta}\cap L{}_{O}^{\delta\prime}=\left\{ \delta\right\} $.
\end{defn}

When we put two pQTSs in parallel, they synchronize on shared actions, and evolve independently on others. 
Since the transitions taken by the two component of the composition are stochastically independent, we
multiply the probabilities when taking shared actions. 

\begin{defn}
\label{def:(Parallel-Composition)}(Parallel composition) Given two
compatible pQTS $\mathcal{A}=\left( S,s_{0},L_{I},L_{O}^{\delta},\Delta\right) $
and $\mathcal{A}'=\left( S',s_{0}',L_{I}',L_{O}^{\delta\prime},\Delta'\right) $,
their \textit{parallel composition} is the tuple
\[
\mathcal{A}\mid\mid\mathcal{A}'=\left( S'',s_{0}'',L_{I}'',L_{O}^{\delta\prime\prime},\Delta''\right) ,
\] where

$S''=S\times S'$,

$s_{0}''=\left(s_{0},s_{0}'\right)$,

$L_{I}''=\left(L_{I}\cup L_{I}'\right)\backslash\left(L_{O}\cup L_{O}^\prime\right)$,

$L_{O}^{\delta\prime\prime}=L_{O}^{\delta}\cup L_{O}^{\delta\prime}$,

$\Delta''=\{\left(\left(s,t\right),\mu\right)\in S''\times\mathit{Distr}\left(L''\times S''\right)\mid$
\[ \mu\left(a,\left(s',t'\right)\right)\equiv\begin{cases}
\mu_{a}\left(a,s'\right)\nu_{a}\left(a,t'\right) & \mbox{if }a\in L\cap L'\mbox{, where }s\overset{\mu_{a},a}{\longrightarrow}_{\mathcal{A}}s'\wedge t\overset{\nu_{a},a}{\longrightarrow}_{\mathcal{A}'}t'\\
\mu_{a}\left(a,s'\right) & \mbox{if }a\in L\backslash L'\mbox{, where }s\overset{\mu_{a},a}{\longrightarrow}_{\mathcal{A}}s'\wedge t=t'\\
\nu_{a}\left(a,t'\right) & \mbox{if }a\in L'\backslash L\mbox{, where }s=s'\wedge t\overset{\nu_{a},a}{\longrightarrow}_{\mathcal{A}'}t'\\
0 & \mbox{otherwise}
\end{cases}\}
\]  where $\mu_{a}\in\mathit{Distr}\left(L,S\right)$ and $\nu_{a}\in\mathit{Distr}\left(L',S'\right)$ respectively.
\end{defn}

Before we parallel compose a test case with a system, we obviously
need to define which outcome of a test case is considered correct, and which is not (i.e., when it fails).

\begin{defn}
(Test case annotation) For a given test $t$ a \textit{test annotation}
is a function

\[
a:\mathit{ctraces}\left(t\right)\longrightarrow\left\{ \mathit{pass},\mathit{fail}\right\} .
\]
A pair $\hat{t}=\left(t,a\right)$ consisting of a test and a test
annotation is called an \textit{annotated test}. The set of all such
$\hat{t}$ is defined as $\hat{T}=\left\{ \left(t_{i},a_{i}\right)_{i\in\mathcal{I}}\right\} $
for some index set $\mathcal{I}$ is called \textit{annotated Test
Suite}. 
If $t$ is a test case for a specification $\mathcal{A}_{s}$ we define
the \pioco test annotation $a_{\mathcal{A}_{s},t}^{\mathit{pioco}}:\mathit{ctraces}\left(t\right)\longrightarrow\left\{ \mathit{pass},\mathit{fail}\right\} $
by
\[
a_{\mathcal{A}_{s},t}^{\mathit{pioco}}\left(\sigma\right)=\begin{cases}
\mathit{fail} & \mbox{if }\exists\sigma_{1}\in\mathit{traces}\left(\mathcal{A}_{s}\right),a!\in L_{O}^{\delta}:\sigma_{1}a!\sqsubseteq\sigma\wedge\sigma_{1}a!\notin\mathit{traces}\left(\mathcal{A}_{s}\right);\\
\mathit{pass} & \mbox{otherwise.}
\end{cases}
\]

\end{defn}

\subsection{Test execution.}
By taking the intersection of all complete traces within a test and all traces of an implementation, we will define the set of all traces that will be executed by an annotated test case.
\begin{defn}
\label{def:(Test-execution)}(Test execution) Let $t$ be a test over
the action signature $\left(L_{I},L_{O}^{\delta}\right)$ and the pQTS $\mathcal{A}_{i}=\left(S,s_{0},L_{I},L_{O}^{\delta},\Delta\right)$. Then we define 

\[
\mathit{exec}_{t}\left(\mathcal{A}_{i}\right)=\mathit{traces}\left(\mathcal{A}_{i}\right)\cap\mathit{ctraces}\left(\hat{t}\right).
\]
\end{defn}

\begin{example}
Consider the specification of a shuffle music player and a derived test for it given in Figure \ref{fig:SpecAndTest}. Assuming we are to test whether or not the following two implementations conform to the specification with respect to \pioco :

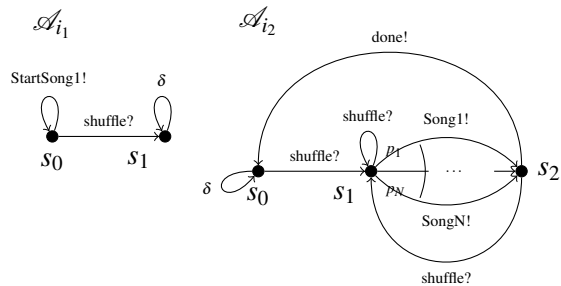
\begin{center}
\begin{tikzpicture}

\node[fill, circle,draw=black, inner sep=1.5, minimum size=0.1, label=below:$s_0$] (init) at (0,0) {};
\node[fill, circle,draw=black, inner sep=1.5, minimum size=0.1, label=225:$s_1$] (select) at (1.5,0) {};

\node (label) at (0,1.5) {$\mathcal{A}_{i_1}$};
\node (label) at (0,-1.95) {};

%% Drawing the transitions

\draw[->] (init) to[out=70, in=120, looseness=25] node[above] {\tiny StartSong1!} (init);
\draw[->] (select) to[out=70, in=120, looseness=25] node[above] {\tiny $\delta$} (select);
\draw[->] (init) to node[above]{\tiny shuffle?} (select);

\end{tikzpicture}
\begin{tikzpicture}

\node[fill, circle,draw=black, inner sep=1.5, minimum size=0.1, label=below:$s_0$] (init) at (0,0) {};
\node[fill, circle,draw=black, inner sep=1.5, minimum size=0.1, label=225:$s_1$, label=45:\tiny$p_1$, label=315:\tiny$p_N$] (select) at (1.5,0) {};
\node[fill, circle,draw=black, inner sep=1.5, minimum size=0.1, label=right:$s_2$] (done) at (3.5,0) {};

\node (arcBegin) at (2.1,0.49) {};
\node (arcEnd) at (2.1,-0.49) {};

\node (label) at (0,2) {$\mathcal{A}_{i_2}$};

\node (afterSelect) at (2.4,0) {};
\node (beforeDone) at (3.0,0) {};
\node (ldots) at (2.6,0) {\tiny$\ldots$};

%% Drawing the transitions
\draw[-] (select) to (afterSelect);
\draw[->] (beforeDone) to (done);

\draw[->] (init) to node[above]{\tiny shuffle?} (select);
\draw[->] (select) to[in=-225, out=45] node[above]{\tiny  Song1!} (done);
\draw[->] (select) to[in=225,  out=-45] node[below]{\tiny SongN!}(done);
\draw[->] (done) to[in=-90,  out=-90, distance=1.5cm] node[below]{\tiny shuffle?}(select);
\draw[->] (done) to[in=90,  out=90, distance=2cm] node[above]{\tiny done!}(init);
\draw[-] (arcBegin) to[in=70, out=290] (arcEnd);

\draw[->] (select) to [in=120, out=70, looseness=25] node[above] {\tiny shuffle?} (select);
\draw[->] (init) to [out=180, in=230, looseness=25] node[left] {\tiny $\delta$} (init);
\end{tikzpicture}
\end{center}
Here $p_{1},\ldots,p_{N}\in\left[0,1\right]$ such that $\sum_{i=1}^{N}p_{i}=1$.
Now when we compose $\mathcal{A}_{i_{1}}$ with $t$ in Figure$\ref{fig:Test}$,
we can clearly see that every complete trace of the parallel system
is annotated with $\mathit{fail}$, as it would also have been the
case for classical \ioco theory. However, if we now also
consider $\mathcal{A}_{i_{2}}$ and compose it with the same test
$t$, every trace of the composed system would be given a $\mathit{pass}$
label if we restricted ourselves to the annotation function and the
output verdict. Note how every trace $\mathit{shuffle?}\cdot\mathit{Song\_}i!$
is given probability $p_{i}$ for $i=1,\ldots,N$. The only
restriction we assumed valid for $p_{1},\ldots,p_{N}$ is that they
sum up to $1$ so a correct distribution for $\mathcal{A}_{i_{2}}$
would be $p_{1}=\frac{N-1}{N}$ and $p_{2}=\ldots=p_{N}=\frac{1}{N^{2}}$.
This, however, should intuitively not be given the verdict $\mathit{pass}$,
since it differs from the uniform distribution given in the specification
$\mathcal{A}_{s}$. 
\end{example}

\subsection{Test evaluation}
In order to give a verdict of whether or not the implementation passed the test (suite), we need to extend the test evaluation process of classical ioco testing with a statistical component. Thus the idea of evaluating probabilistic systems becomes two folded. On the one hand, we want that no unexpected output (or unexpected quiescence) ever occurs during the execution. On the other hand, we want the observed frequencies of the SUT to conform in some way to the probabilities described in the specification. Thus the SUT will pass the test suite only if it passes both criteria. We will do this by augmenting classical \ioco theory with zero hypothesis testing, which will be discussed in the following.

To conduct an experiment, we need to define a length $k\in\mathbb{N}$ and a width $m\in\mathbb{N}$ first. This refers to how long the traces we want to record should be and how many times we reset the machine. This will give us traces $\sigma_1,\ldots,\sigma_m\in L^k$, which we call a \textit{sample}. Additionally, we assume that the implementation is governed by an underlying trace distribution $H$ in every run, thus running the machine $m$ times, gives us a sequence of possibly $m$ different trace distributions $\vec{H}=H_1,\ldots,H_m$. So in every run the implementation makes two choices: 1) It chooses the trace distribution $H$ and 2) $H$ chooses a trace $\sigma$ to execute. Consequently that means that once a trace distribution $H_i$ is chosen, it is solely responsible for the trace $\sigma_i$. Thus for $i\neq j$ the choice of $\sigma_i$ is independent from the choice of $\sigma_j$.

Our statistical analysis is build upon the frequencies of traces occurring in a sample $O$. Thus the \textit{frequency function} will be defined as 
\[
\mathit{freq}\left(O\right)\left(\sigma\right)=\frac{\left|\left\{ i\in,\left\{ 1,\ldots,m\right\} |\sigma_{i}=\sigma\right\} \right|}{m}.
\]
Note that although every run is governed by possibly different trace distributions, we can still derive useful information from the frequency function. For fixed $k,m\in\mathbb{N}$ and $\vec{H}$, the sample $O$ can be treated as a Bernoulli experiment of length $m$, where success occurs in position $i=1,\ldots m$, if $\sigma=\sigma_i$. The success probability is then given by $P_{H_i}\left(\sigma\right)$. So for given $\vec{H}$, the expected value for $\sigma$ is given by $\mathbb{E}^{\vec{H}}_\sigma=\frac{1}{m}\sum_{i=1}^m P_{H_i}\left(\sigma\right)$. Note that this expected value $\mathbb{E}^{\vec{H}}$ is the expected distribution over $L^k$ if we assume it is based on the $m$ trace distributions $\vec{H}$.

In order to apply zero hypothesis testing and compare an observed distribution with $\mathbb{E}^{\vec{H}}$, we will use the notion of metric spaces. This will enable us to measure deviation of two distributions. We will use the metric space $\left(L^{k},\mathit{dist}\right)$, where $dist$ is the euclidean distance of two distributions defined as $\mathit{dist}\left(\mu,\nu\right)=\sqrt{\sum_{\sigma\in L^k}\left|\mu\left(\sigma\right)-\nu\left(\sigma\right)\right|^2}$.

Now that we have a measure of deviation, we can say that a sample $O$ is accepted if $\mathit{freq}\left(O\right)$ lies in some distance $r$ of the expected value $\mathbb{E}^{\vec{H}}$, or equivalently if $\mathit{freq}\left(O\right)$ is contained in the closed ball $B_r\left(\mathbb{E}^{\vec{H}}\right)=\left\{\nu\in\mathit{Distr}\left(L^{k}\right)\mid \mathit{dist}\left(\nu,\mathbb{E}^{\vec{H}}\right)\leq r\right\}$. Then the set $\mathit{freq}^{-1}\left(B_r\left(\mathbb{E}^{\vec{H}}\right)\right)$ summarizes all samples that deviate at most $r$ from the expected value. 

An inherent problem of hypothesis testing are the type 1 and type 2 errors, i.e. the probability of falsely accepting the hypothesis or falsely rejecting it. This problem is established in our framework by the choice of a level of significance $\alpha\in\left[0,1\right]$ and connected with it, the choice of radius $r$ for the ball mentioned above. So for a given level of significance $\alpha$ the following choice of the radius will in some sense minimize the probability of false acceptance of an erroneous sample and of false rejection of a valid sample (i.e., at most $\alpha$).
\[
\bar{r}:=\inf\left\{r\mid P_{\vec{H}}\left(\mathit{freq}^{-1}\left(B_r\left(\mathbb{E}^{\vec{H}}\right)\right)\right)>1-\alpha\right\}.
\]
Thus assuming we have $m$ different underlying trace distributions, we can determine when an observed sample seems reasonable and is declared valid. Unifying over all sets of such $\vec{H}$, we will define the total set of acceptable outcomes, called \textit{Observations}.
\begin{defn} The \textit{acceptable outcomes} of $\vec{H}$ with significance level $\alpha\in\left[0,1\right]$ are given by the set of samples of length $k\in\mathbb{N}$ and width $m\in\mathbb{N}$, defined as
\[
\mathit{Obs}\left(\vec{H},\alpha\right):=\mathit{freq}^{-1}\left(B_{\bar{r}}\left(\mathbb{E}^{\vec{H}}\right)\right)=\left\{O\in \left(L^k\right)^m\mid \mathit{dist}\left(\mathit{freq}\left(O\right),\mathbb{E}^{\vec{H}}\right)\leq\bar{r}\right\}.
\]
The set of \textit{observations} of $\mathcal{A}$ with significance level $\alpha\in\left[0,1\right]$ is given by
\[
\mathit{Obs}\left(\mathcal{A},\alpha\right)=\bigcup_{\vec{H}\in\mathit{trd}\left(\mathcal{A},k\right)^m}\mathit{Obs}\left(\vec{H},\alpha\right).
\]
\end{defn}

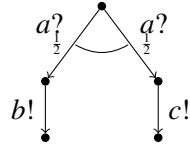
\begin{figure}
\begin{center}
\begin{tikzpicture} 
\node[fill, circle,draw=black, inner sep=1, minimum size=0.1] (initImp) at (1,1) {} ;
\node[fill, circle,draw=black, inner sep=1, minimum size=0.1] (leftImp) at (0.25,0) {};
\node[fill, circle,draw=black, inner sep=1, minimum size=0.1] (rightImp) at (1.75,0) {};
\node[fill, circle,draw=black, inner sep=1, minimum size=0.1] (leftUnd) at (0.25,-0.75) {};
\node[fill, circle,draw=black, inner sep=1, minimum size=0.1] (rightUnd) at (1.75,-0.75) {};
\draw[->] (initImp) to node[inner sep=0,label=135:$a?$, label=left:\tiny $\frac{1}{2}$](middle1){} (leftImp);
\draw[->] (initImp) to node[inner sep=0,label=45:$a?$, label=right:\tiny $\frac{1}{2}$](middle2){} (rightImp);
\draw (middle1) to[bend right] (middle2);

\draw[->] (leftImp) to node[inner sep=0, label=left:$b!$](middle3) {} (leftUnd);
\draw[->] (rightImp) to node[inner sep=0, label=right:$c!$](middle4) {} (rightUnd);

\end{tikzpicture}\caption{A probabilistic automaton representing a fair coin.}\label{fig(fairCoin)}\end{center}\end{figure}
\begin{example} Assume that the wanted level of significance is given by $\alpha=0.05$
  and consider the probabilistic automaton in Figure \ref{fig(fairCoin)} representing the toss of a fair coin. Furthermore assume that we are given two samples of depth $k=2$
  and width $m=100$.
  
  To sample this case, assume $E$ is the adversary that assigns probability equal to $1$
  to the unique outgoing transition (if there is one) and probability $1$ to halting, in case there is no outgoing transition. We take $H=\mathit{trd}\left(E\right)$ and can see, that then $\mu_H\left(a?b!\right)=\mu_H\left(a?c!\right)=\frac{1}{2}$ and $\mu_H\left(\sigma\right)=0$ for all other sequences $\sigma$. We define $H^{100}=\left(H_{1},\ldots,H_{100}\right)$, where $H_1=\ldots=H_{100}=H$. 
  As we can see, we have $\mathbb{E}^{H^{100}}=\mu_H$. Since $\mu_H$ only assigns positive probability to $a?b!$ and $a?c!$, we get $P_{H^{100}}\left(B_r\left(\mu_H\right)\right)=\left(\left\{O|\frac{1}{2}-r\leq\mathit{freq}\left(O\right)\left(a?b!\right)\leq\frac{1}{2}+r\right\}\right)$. One can show that the smallest ball, where this probability is greater or equal than  $0.95$ is given by the ball of radius $\bar{r}=\frac{1}{10}$.

  Thus a sample $O_1$, which consists of $42$ times $a?b!$ and $58$ times $a?c!$ is an observation, and a sample $O_2$, which consists of $38$ times $a?b!$ and $62$ times $a?c!$ is not.
\end{example}

Thus we can finally define a verdict function, that assigns \textit{pass} when a test case never finds erroneous behaviour (i.e. wrong output or wrong probabilistic behaviour).
\begin{defn}
(Output verdict) Let $\left(L_{I},L_{O}^{\delta}\right)$ be an action
signature and $\hat{t}=\left(t,a\right)$ an annotated test case over
$\left(L_{I},L_{O}^{\delta}\right)$. The \textit{output verdict function}
for $\hat{t}$ is the function $v_{\hat{t}}:pQTS\rightarrow\left\{ \mathit{pass},\mathit{fail}\right\} $,
given for any pQTS $\mathcal{A}_{i}$
\[
v_{\hat{t}}\left(\mathcal{A}_{i}\right)=\begin{cases}
\mathit{pass} & \mbox{if }\forall\sigma\in\mathit{exec}_{t}\left(\mathcal{A}_{i}\right):a\left(\sigma\right)=\mathit{pass}\\
\mathit{fail} & \mbox{otherwise}
\end{cases}.
\]
(Statistical verdict) Additionally let $\alpha\in\left[0,1\right]$
 and $k,m\in\mathbb{N}$ and $O\in\mathit{Obs}\left(\mathcal{A}_{i}||\hat{t},\alpha\right)\subseteq\left(L^{k}\right)^{m}$,
then the \textit{statistical verdict function} is given by 
\[
v_{\hat{t}}^{\alpha}\left(\mathcal{A}_{i}\right)=\begin{cases}
\mathit{pass} & \mbox{if }O\in\mathit{Obs}\left(\mathcal{A}_{s},\alpha\right)\\
\mathit{fail} & \mbox{otherwise }
\end{cases}.
\]
(Verdict function) For any given $\mathcal{A}_{i}$, we assign the
\textit{verdict}
\[
V_{\hat{t}}^{\alpha}\left(\mathcal{A}_{i}\right)=\begin{cases}
\mathit{pass} & \mbox{if }v_{\hat{t}}\left(\mathcal{A}_{i}\right)=v_{\hat{t}}^{\alpha}\left(\mathcal{A}_{i}\right)=\mathit{pass}\\
\mathit{fail} & \mbox{otherwise}
\end{cases}.
\]
We extend $V_{\hat{t}}^{\alpha}$ to a function $V_{\hat{T}}^{\alpha}:pQTS\rightarrow\left\{ \mathit{pass},\mathit{fail}\right\} $,
which assigns verdicts to a pQTS based on a given annotated test suite
by $V_{\hat{T}}^{\alpha}\left(\mathcal{A}_{i}\right)=\mathit{pass}$
if for all $\hat{t}\in\hat{T}$ and $V_{\hat{T}}^{\alpha}\left(\mathcal{A}_{i}\right)=\mathit{fail}$
otherwise. 
\end{defn}

\section{Conclusion and Future Work}\label{sec(FutureWork)}

We introduced the core of a probabilistic test theory by extending classical \ioco theory. We defined the conformance relation \pioco for probabilistic quiescent transition systems, and proved several characteristic properties. In particular, we showed that \pioco is a conservative extension of \ioco. Second, we have provided definitions of a test case, test execution and test evaluation. Here, test execution is crucial, since it needs to assess whether the observed behaviour respects the probabilities in the specification pQTS. Following \cite{CSV07}, we have used statistical hypothesis testing here.

Being a first step, there is ample future work to be carried out. First, it is important to establish the correctness of the testing framework, by showing the soundness and completeness. Second, we would like to implement our framework in the MBT testing framework JTorX, and test realistic applications. Also, we would like to extend our theory to handle $\tau$-transitions. Finally, we think that tests themselves should be probabilistic, in particular since many MBT tools in practice do already choose their next action probabilistically.

\nocite{*}
\bibliographystyle{eptcs}
\bibliography{biblio.bib}

\newpage
\begin{appendix}
\section*{Appendix}

Below, we present the proofs of our theorems.

\section*{Proofs}
\printproofs
\end{appendix}

\end{document}